\def\year{2023}\relax
\documentclass[letterpaper]{article} 
\usepackage{aaai22}  
\usepackage{times}  
\usepackage{helvet} 
\usepackage{courier}  
\usepackage[hyphens]{url}  
\usepackage{graphicx} 
\usepackage{natbib}  
\usepackage{caption} 

\usepackage{amsmath}
\usepackage{amssymb}
\usepackage{makecell}
\usepackage{float}
\usepackage{subcaption}
\usepackage{marginnote}

\let\oldmarginnote\marginnote
\renewcommand{\marginnote}[1]
     {\begingroup\docolaction{\reversemarginpar}{}{}\oldmarginnote{#1}\endgroup}

\usepackage{mathptmx}
\usepackage[english]{babel}
\usepackage{csquotes}
\usepackage{xcolor}
\usepackage{verbatim}
\usepackage{booktabs}
\usepackage{breqn}

\newcommand{\mycomment}[1]{}
\title{Quotatives Indicate Decline in Objectivity in U.S. Political News}

\author {
    Tiancheng Hu\textsuperscript{\rm 1},
    Manoel Horta Ribeiro\textsuperscript{\rm 2},
    Robert West\textsuperscript{\rm 2},
     Andreas Spitz\textsuperscript{\rm 3}
}
\affiliations {
    \textsuperscript{\rm 1} ETH Zürich, Zürich, Switzerland\\
    \textsuperscript{\rm 2} EPFL, Lausanne, Switzerland\\
    \textsuperscript{\rm 3} University of Konstanz, Konstanz, Germany\\
    tiancheng.hu@alumni.ethz.ch, manoel.hortaribeiro@epfl.ch, 
    robert.west@epfl.ch,
    andreas.spitz@uni-konstanz.de
}

\begin{document}

\maketitle

\begin{abstract}
According to journalistic standards, direct quotes should be attributed to sources with objective quotatives such as ``said'' and ``told,'' since nonobjective quotatives, e.g., ``argued'' and ``insisted,'' would influence the readers' perception of the quote and the quoted person.
In this paper, we analyze the adherence to this journalistic norm to study trends in objectivity in political news across U.S. outlets of different ideological leanings. 
We ask: 
1)~How has the usage of nonobjective quotatives evolved? 
2)~How do news outlets use nonobjective quotatives when covering politicians of different parties? 
To answer these questions, we developed a dependency-parsing-based method to extract quotatives and applied it to Quotebank, a web-scale corpus of attributed quotes, obtaining nearly 7 million quotes, each enriched with the quoted speaker's political party and the ideological leaning of the outlet that published the quote.
We find that, while partisan outlets are the ones that most often use nonobjective quotatives, between 2013 and 2020, the outlets that increased their usage of nonobjective quotatives the most were ``moderate'' centrist news outlets (around 0.6 percentage points, or 20\% in relative percentage over seven years).
Further, we find that outlets use nonobjective quotatives more often when quoting politicians of the opposing ideology (e.g., left-leaning outlets quoting Republicans) and that this ``quotative bias'' is rising at a swift pace, increasing up to 0.5 percentage points, or 25\% in relative percentage, per year.
These findings suggest an overall decline in journalistic objectivity in U.S. political news. 
\end{abstract}

\section{Introduction}\label{sec:intro}

Journalistic objectivity is the notion that news should contain accurate information and not convey the personal opinions or emotions of the writer~\cite{ryan2001journalistic,calcutt2011journalism}. 
Historically, objectivity emerged alongside the conception of journalism as a profession~\cite{schudson1978discovering} and has shaped many of the practices and norms in modern journalism~\cite{boudana2011definition}. 
In the context of U.S. politics, with its two major political parties, this can also be interpreted as ``equal treatment'' of both parties~\cite{d2000media}. 
Bias in the news could affect public opinion~\cite{o1999irish,kahn2002slant} and lead to changes in voting behavior~\cite{dellavigna2007fox,bernhardt2008political}.

``Absolute'' objectivity has been criticized as unattainable, as structural biases would creep into news production~\cite{badnews}, or even as harmful, as the excessive balance of viewpoints could create an illusion of credibility for dubious or unsupported positions~\cite{dixon2013heightening}.
However, amidst the fragmented media ecosystem that emerged from the digitization of news outlets and the algorithmic serving of content~\cite{thurman2019algorithms}, journalism scholars have argued that objectivity has become ever more important to consumers of journalism~\cite{boudana2011definition,mcnair2017after}. 
This opinion is also held by the public worldwide, who, as of 2018, overwhelmingly agree that news media should be unbiased in its coverage of politics~\cite{mitchell2018publics}.

One of the concrete ways in which journalists have sought to report the news objectively is through the usage of direct quotes~\cite{brooks2007news,stenvall2008emotions}.
Since journalists almost never directly observe the events they report, using quotes lends them more reliability and factuality than their own words~\cite{van2013news}. 
Furthermore, direct quotes would let people ``speak for themselves,'' following one of the golden rules of journalism~\cite{thenewsmanual}.
However, even when using a direct quote, journalistic objectivity can still be compromised by the use of certain quotatives that relay the emotions of reporters to readers~\cite{mencher1997news} or the attempt to describe the speaker's state of mind~\cite{gidengil2003talking}. 
For example, in the direct quote 
\begin{displayquote}
    \textit{``New York is not afraid of terrorists,''} \textbf{boasted} \underline{Rep. Jerrold} \underline{Nadler}, a Democrat representing Manhattan,
\end{displayquote}
the quotative (\textbf{boasted}) carries an illocutionary force from the reporter that influences how the reader perceives the quote itself, possibly distorting its original meaning~\cite{caldas1992reporting}. 
Objective quotatives, like ``say'' or ``tell,'' on the other hand, are considered neutral, as they imply little about the presumed intent or the fashion in which the quote was uttered~\cite{sonoda1997subject,languageofnewsmedia}.

Recent years were marked by increased political polarization~\cite{abramowitz2008polarization}, mistrust in media~\cite{brenan2022americans}, increased negative tone by politicians \cite{DBLP:journals/corr/abs-2207-08112}, and the perception that the public debate around politics has become less respectful and less fact-based~\cite{doherty2019public}. 
Solutions to these issues are complex, but analyzing the bias and the departure from journalistic objectivity in political news coverage can help inform new practices and interventions that seek to improve the political news media ecosystem. 
Quotatives, in this context, are a powerful instrument to measure bias.
Studying how journalists deviate from the standard usage of quotatives -- e.g., ``say'' and ``tell'' \cite{apstylebook} -- allows researchers to quantitatively assess adherence to journalistic objectivity~\cite{lee2017verb} and reveal biases in journalistic coverage of politics~\cite{gidengil2003talking}. 

\subsubsection{Present work.}
This paper analyzes quotatives to study objectivity and media bias in political journalism. We ask:
\begin{itemize}
    \item \textbf{RQ1}  How has the usage of nonobjective quotatives evolved in U.S. political journalism?
    \item \textbf{RQ2}  How do news outlets use nonobjective quotatives when covering politicians of different parties? 
\end{itemize}
To answer these research questions, we developed a methodology to extract quotatives from a large-scale news corpus. 
We then performed a comprehensive study on how (and which) quotatives are used in direct quotes from U.S. politicians between 2013 and 2020, leveraging a large dataset of quotes from English-language media linked with relevant speaker metadata~\cite{vaucher2021quotebank} and enriched with the political leanings of different U.S. outlets.
By counting the usage of nonobjective quotatives like ``shout'' or ``assert,'' we analyze how U.S. news outlets of different political inclinations adhere to basic journalistic objectivity principles and how this adherence has evolved.
Further, analyzing how outlets of different political inclinations use quotatives to talk about politicians of different parties, we study the evolution of quotative bias in news outlets.

\subsubsection{Summary of findings.}
We find that the usage of nonobjective quotatives varies across different outlet categories. Overall, the more ideologically extreme an outlet is, the more nonobjective quotatives it uses. 
However, we also find that centrist outlets are experiencing a significant increase in the usage of nonobjective quotatives over the last years (about 0.6 percentage points, or 20\% in relative percentage), suggesting that they may be ``catching up'' to the more biased outlets, which are not experiencing such significant increases (\textbf{RQ1}). 
We also find evidence of ``quotative bias,'' i.e., outlets tend to use nonobjective quotatives, especially when referring to politicians of opposing ideology.
For instance, left and right-leaning outlets use nonobjective quotatives up to 2\% more often when referring to Republican and Democrat politicians, respectively (\textbf{RQ2}).
Last, we find that this quotative bias is increasing at a swift pace, increasing as much as 0.5 percentage points per year in absolute percentage, or 25\% in relative percentage, for left-leaning outlets, suggesting a rapid increase in polarization (\textbf{RQ1} and \textbf{RQ2}). 

\subsubsection{Implications.}
Our findings indicate a decline in journalistic objectivity in U.S.  political news, particularly from centrist outlets. 
This suggests that centrist outlets may play a role in the increasingly less respectful and fact-based debate around politics~\cite{doherty2019public}.
Further, we also find evidence of an increasing quotative bias, which could further erode trust in the media~\cite{brenan2022americans}. 

\section{Background and Related Work}\label{sec:background}
When a quote occurs in the news, three elements are typically involved: the source, i.e., the speaker who uttered the quote (\underline{underlined} in the examples); 
the quoted content itself (in \textit{italic}); and the quotative that introduces the quote (also known as a cue, reporting verb, speech verb, or attribution verb; in \textbf{bold}). 
Quotes can be classified as either direct, indirect, mixed, or pure~\cite{cappelen1997varieties}, where only the former three types are typically of concern to journalists. We give an example of a direct quote in the introduction and of mixed and indirect quotes below.

\begin{displayquote}

\textbf{Indirect quote:}
\underline{Sen. Ron Wyden} of Oregon, the chairman of the Senate Finance Committee, \textbf{indicated} that \textit{in 2019, about 100 to 125 corporations reported financial statement income greater than 1B USD.}

\textbf{Mixed quote:}
\underline{Catsimatidis} \textbf{said} he'd serve for 99 cents \textit{``because I'm a grocer.''}
\end{displayquote}

In direct and mixed quotes, a pair of quotation marks are used, and we can infer that the speaker uttered the quoted words, whereas indirect quotes may paraphrase the speaker's words. Therefore, journalists have the most freedom in word choice in indirect quotes, as they can, to some extent, rewrite what the speaker said. 
In contrast, in direct quotes, journalistic norms require them to report the quoted words verbatim~\cite{harry2014journalistic}.

In our work, we focus on direct quotes for the following reasons:  Quotebank does not contain indirect quotes~\cite{vaucher2021quotebank}; automatically extracting the quotative in mixed quotes is technically challenging and, in some instances, impossible as there is no quotative, e.g.,
\underline{John} will not help as he has \textit{``done more for this house than all of us combined''.}

\vspace{1.5mm}
\noindent
\textbf{Measuring media bias.}
\label{sec:meas_bias}
Previous work has studied media biases:  how journalists' and editors' personal opinions, beliefs, and financial incentives shape what is considered newsworthy~\cite{mccombs1972agenda} and how issues are covered~\cite{iyengar1994anyone}.
Scholars argue that partisan media bias can harm democracy by distorting citizens' political knowledge and increasing polarization~\cite{bernhardt2008political,boudana2011definition,mcnair2017after}. 
Thus, measuring media bias is the first step to improving our information ecosystem~\cite{watts}.

Early studies in media bias required extensive manual annotation. 
For instance, \citet{kobre1953florida} studied how the press in Florida covered the U.S. 1952 presidential campaign by coding the number of inches of text given to each party, the position of pictures, etc., across hundreds of newspapers. 
However, in recent times and with the digitization of news, various methods have been developed to \textit{automatically} measure media bias~\cite{hamborg2019automated}.
Some of these methods are \textit{audience based}, measuring how segregated news consumers are across outlets, e.g., \citet{zhou2011classifying} use votes on Digg, a social news aggregator, to classify political articles.
Others are \textit{content based,} quantifying media bias by analyzing published content directly. 
For instance, \citet{gentzkow2010drives} measured bias using the frequency at which outlets reproduce partisan phrases in congressional speeches.

According to \citet{budak2016fair}, both content and audience-based approaches suffer from distinctive limitations. 
On the one hand, audience-based approaches do not scale beyond outlets for which detailed readership information can be obtained. 
On the other hand, content-based approaches struggle to generalize well across different types of news and outlets, e.g., methods that try to match politicians' speeches to news only apply to a minority of news articles, limiting the scope of the results obtained. 

\vspace{1.5mm}
\noindent
\textbf{Quotatives and bias.}
\label{sec:quot_bias}
Quotatives can impact how readers perceive a news story and the involved speakers~\cite{geis1987language,just1999voice}. 
For instance, \citet{cole1974powerful} carried out an experiment in which they changed ``objective'' quotatives like ``said'' for stronger verbs like ``argued'' or ``insisted'' and asked participants to rate stories across a variety of criteria. 
They found that in the modified versions, stories were perceived as more exciting and less objective, and speakers were perceived as more rash.
Through quotatives, journalists can ``paint reports on speech with any brush they like''~\cite{geis1987language}, which would not only reveal the beliefs and preferences of the writer~\cite{gidengil2003talking} but also subtly influence the reader~\cite{cole1974powerful}.
In this context, quotatives have been used to measure political bias, sometimes referred to as ``attribution bias.'' 
This line of work dates from the 1960s when \citet{merrill1965time} studied how Time magazine used quotatives (among other things) when referring to U.S. Presidents Kennedy, Truman, and Eisenhower. 
More recently, \citet{gidengil2003talking} analyzed differences in quotative usage between male and female party leaders on Canadian television, finding that female leaders' speech was reported with more negative and aggressive quotatives. 
With a similar methodology, \citet{lee2017verb} studied differences in nonobjective quotatives between offline and online newspapers, finding the former to adhere better to journalistic standards.

\vspace{1.5mm}
\noindent
\textbf{Quote attribution and analysis.}
\label{ref:quots}
Previous work in natural language processing has studied the problem of \textit{quote attribution} (see \citet{vaucher2021quotebank} for a review), an important task in understanding dialogue structure and developing better conversational agents. 
For each quote, the goal is to extract the speaker of the quote, either at the mention or entity level.
This task is challenging as the speaker could be mentioned implicitly or require anaphora resolution. 
The task can be further combined with entity linking to extract unique IDs of speakers~\cite{culjak-etal-2022-strong}. Most prior work, however, has not dealt with the problem of quotative extraction.

Nonetheless, several datasets annotated for attributional relationship exist~\cite{pareti2012database,pareti-2016-parc3,newell2018attribution} that could be considered in this context. 
These datasets contain labels for the content, source, and cue for each attributional relationship. They can be viewed as an extension of  The Penn Discourse TreeBank 2.0 \cite{prasad-etal-2008-penn} that provides annotation of discourse relations and argument structures. 
While these datasets can potentially be used as resources for training a supervised model for quotative extraction, the attributional relationship they considered is much broader than quotation and thus not suitable for our study. 

Existing work has also analyzed quotes from different perspectives. \citet{niculae2015quotus} found a systematic pattern in the outlets' quoting behavior when covering the exact same event. 
\citet{lazaridou2017identifying} found that a machine learning classifier could reliably predict one of two news outlets based solely on the quotes they report, demonstrating media bias. 
\citet{tan2018you} showed a declining trend of bipartisan quote coverage with a bipartite graph of media outlets and the sentences they quoted. 
\citet{DBLP:journals/corr/abs-2207-08112} analyzed the quotes of U.S politicians between 2008 and 2020 and found a decrease in negativity during Obama's tenure and a sudden increase starting from Trump's presidential primary campaign in 2015. 

\vspace{1.5mm}
\noindent
\textbf{Relationship with prior work.}
In this paper, we set out to investigate how the usage of nonobjective quotatives evolved in U.S. political journalism (\textbf{RQ1}) and how it is modulated by media biases (\textbf{RQ2}).
We do so by using dependency parsing to extract quotatives from a large dataset (see Sec.~\ref{sec:methods}).
Our method is related to \textit{quote attribution}, a problem widely studied in natural language processing, with the key difference that previous methods aim to attribute  quotes to speakers instead of finding the quotative used.
Further, our approach is similar to previous work that derives automated media bias measurements~\cite{budak2016fair}. However, in contrast to previous work, we automate the measurement of \emph{quotative} bias instead of relying on traditional manual annotation \cite{cole1974powerful}.
Due to the scalability of our approach, we obtain results that help further understand the political news ecosystem (see Sec.~\ref{sec:discussion}). Namely, while previous work often attributes the decrease in journalistic objectivity to the rise of partisan media~\cite{mcnair2017after}, we find that centrist outlets in our dataset have systematically departed from journalistic standards.

\section{Materials and Methods}\label{sec:methods}
\subsection{Data and Data Processing}

To study quotative usage across various news outlets, we use the Quotebank dataset~\cite{vaucher2021quotebank}, a web-scale corpus of quotes. 
Quotebank contains over 235 million unique quotes, extracted from 196 million English news articles from 377 thousand web domains between September 2008 and April 2020.
We additionally obtain a list of current and former U.S. politicians with their party affiliations from Wikidata, in the same fashion as  \citet{DBLP:journals/corr/abs-2207-08112}. 
We filter Quotebank to consider the period containing the best-quality speaker attributions (May 2013 to 2020) and retain only quotes from politicians on this list.
In cases where quotes are attributed to more than one speaker in Quotebank (which happens to 8.13\% of speakers in 12.25\% of the quotes), we heuristically attribute the quote to the speaker with the alphanumerically smallest Wikidata identifier.
We validate speaker attribution in our filtered dataset on a manually annotated sample of 100 quotes and find that combining the speaker names provided by Quotebank with this heuristic yields 86\% accuracy in identifying the correct ID.

\begin{figure}
    \begin{center}
    \includegraphics[scale=0.39]{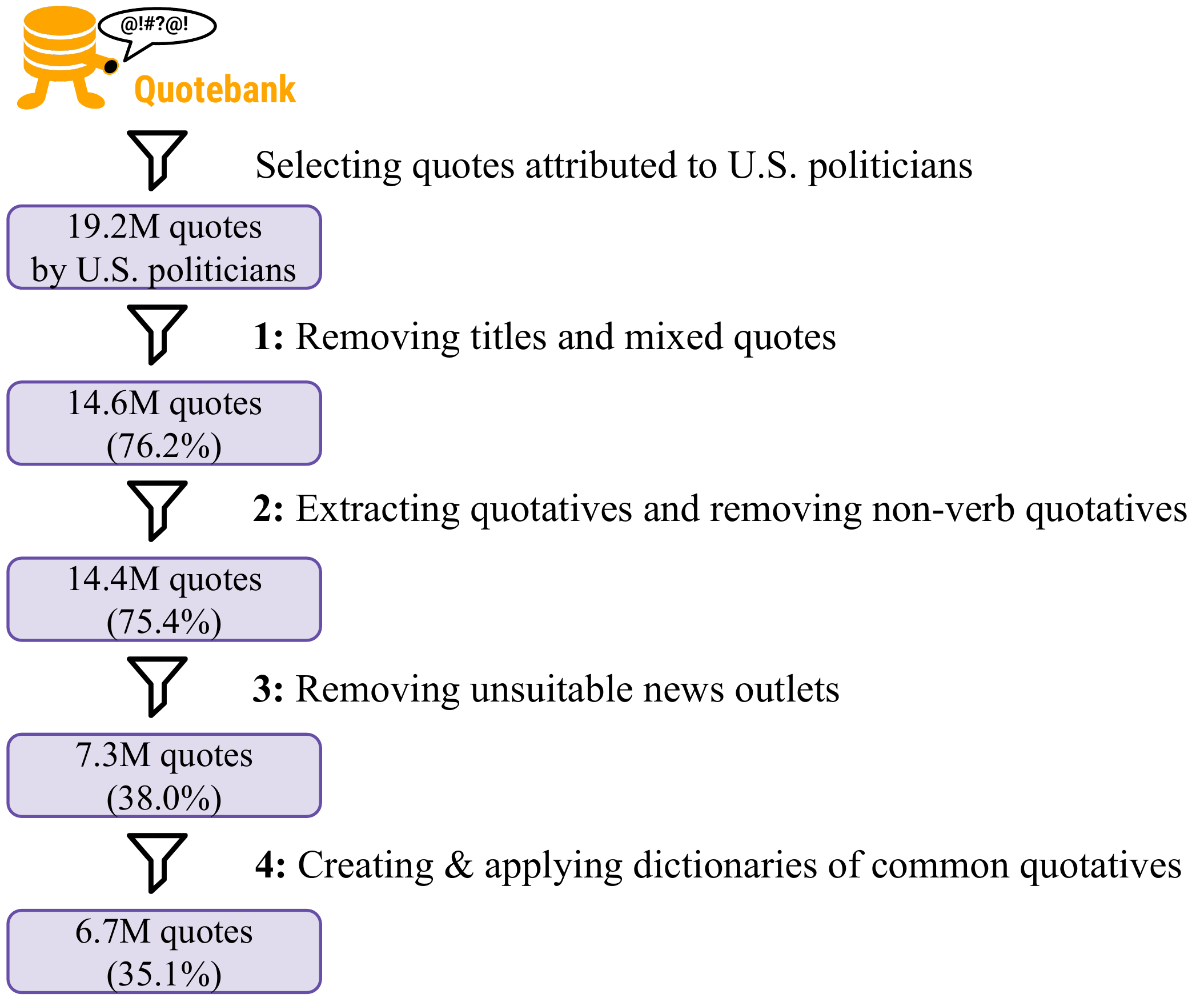} 
    \end{center}
    \caption{Data processing pipeline. We outline the key steps in our data processing pipeline and the percentage of retained data after filtering. }
    \label{fig:data_preprocessing}
\end{figure}

To ensure the validity of our findings, we preprocess Quotebank as depicted in Figure~\ref{fig:data_preprocessing}. 
We 
1) use heuristics to retain only direct quotes;
2) extract quotatives and remove quotes without quotatives in the verb form;
3) filter quotes, keeping only those from U.S-based outlets with human-verified bias ratings; and
4) create dictionaries of common quotatives, removing quotes with rare quotative verbs for which quotative extraction performs poorly.
We detail each of these steps in the following paragraphs.

\vspace{0.5mm}
\noindent
\textbf{Step 1: removing titles and mixed quotes.}
To remove titles from the dataset that are erroneously recognized as quotes (e.g., movies), we apply a filter using the percentage of words in a quote whose first letter is capitalized. 
Afterward, to remove mixed quotes, we employ a sentence recognition filter that combines constituency parsing%
\footnote{Constituency parsing breaks down sentences into phrases and identifies their grammatical roles,
e.g., in  ``I eat a big apple,'' ``a big apple'' is a noun phrase. See \citet{jurafsky2022speech} for details.
}
and dependency parsing%
\footnote{Dependency parsing extracts dependency relationships between words, with verbs typically being in the structural center, e.g., in ``I eat a big apple,'' ``big'' is an adjectival modifier of ``apple.'' See \citet{kubler2009dependency} for details.}.
We retain only quotes that can be parsed as a full sentence at the root level by constituency parsing and contain a subject and a predicate (root) in dependency parsing.
These heuristics improve data quality (e.g., extracted quotatives in Step 2 are much more accurate) while retaining 76.2\% of the dataset.

\vspace{0.5mm}
\noindent
\textbf{Step 2: extracting quotatives and removing non-verb quotatives.} 
In the next step, we adopt a three-stage approach to extract the quotative from each quote using dependency parsing. 
First, we run dependency parsing and acquire a distribution of quotatives from the root node of each parsed quote. Second, we add a condition to ensure that in cases where one verb is identified as the root and another verb exists in a parallel node%
\footnote{\texttt{csubj}, \texttt{ccomp}, \texttt{xcomp}, \texttt{advcl}, \texttt{acl}, \texttt{parataxis}, \texttt{conj}, \texttt{cc}, \texttt{relcl}, 
see \url{https://universaldependencies.org/en/dep/}}, we choose the verb with the higher probability as the quotative (according to the distribution of verbs extracted in the first stage).
Finally, we take the lemma of each extracted quotative and remove quotatives that are not in verb form. 
After this step, we retain around 75.4\% of the original data.

\begin{table*}[tb]
\footnotesize
\begin{center}
\begin{tabular}{lrlrrrrrr}\\\toprule
 \textbf{Quotative} & \textbf{\%} & \textbf{Speaker} & \textbf{\%}  &\textbf{\% Left}&\textbf{\% Left-Center}&\textbf{\%Least-Biased}&\textbf{\%Right-Center}&\textbf{\%Right}\\
\midrule
    say& 72.58 & Donald Trump& 15.64 &19.51&15.59&15.18&13.85&16.93  \\ 
    tell& 8.82  &Barack Obama& 3.50&4.36& 3.90 &2.60&3.39&4.36  \\ 
    write&  3.82 & Hillary Clinton&  1.44 & 2.31&1.47&0.98&1.28&2.35\\ 
    tweet&  2.97 & Nancy Pelosi&  1.35 & 1.26& 1.16& 1.43& 1.26&1.89\\  
    add&  2.56 & Bernie Sanders&  1.24&2.67&1.17&1.00&0.99&1.60\\  
    ask&  1.29 & Joe Biden& 1.23&1.60&1.16&1.13&0.94&1.89\\  
    continue& 0.72 & Mitch McConnell& 1.04 &1.03&0.98&1.11&0.99&1.13\\  
    respond& 0.59 & Chuck Schumer& 0.90&0.73&0.85&0.98&0.85&1.03 \\  
    declare& 0.59 & Lindsey Graham& 0.88&1.13&0.78&0.83 &0.78&1.30\\  
    state& 0.45 & Elizabeth Warren& 0.86&1.43&0.83&0.77&0.69&1.08\\  
\bottomrule
\end{tabular}
\caption{Statistics on top speakers and quotatives.
Frequency of the top quotatives and speakers in the entire dataset after filtering.
We also include the speaker frequency in each outlet category for reference. 
Among the listed quotatives, only ``declare'' is nonobjective.}

\label{tab:most_frenquent_quotative}
 \end{center}
\end{table*}
\begin{table*}[tb]
\footnotesize
\begin{center}
\begin{tabular}{lrlrlrlrlr}\\
\toprule
\textbf{Left} & \textbf{\%} & \textbf{Left-Center} & \textbf{\%} & \textbf{Least-Biased} & \textbf{\%} & \textbf{Right-Center} & \textbf{\%} & \textbf{Right} & \textbf{\%}\\
\multicolumn{2}{l}{$n=80$} & \multicolumn{2}{l}{$n=249$} & \multicolumn{2}{l}{$n=467$} & \multicolumn{2}{l}{$n=142$} & \multicolumn{2}{l}{$n=51$} \\
\multicolumn{2}{l}{$k=0.46$M} & \multicolumn{2}{l}{$k=2.45$M} & \multicolumn{2}{l}{$k=2.12$M} & \multicolumn{2}{l}{$k=0.75$M} & \multicolumn{2}{l}{$k=0.94$M} \\
\midrule     
CNN          & 20.17 & Yahoo    & 6.15 & The Hill  & 6.41 & Washington Times       & 11.50 & Breitbart       & 17.68 \\ 
Raw Story     & 5.17 & MSN      & 5.15 & Roll Call & 1.66 & NW AR Democrat-Gaz. & 3.42 & Fox News        & 11.77 \\ 
Salon         & 5.16 & SFGATE   & 3.44 & KVIA-TV   & 1.38 & AR Democrat-Gaz.    & 3.10 & Wash. Examiner  & 11.62 \\ 
The Week      & 5.02 & WaPo     & 3.28 & UPI       & 1.29 & Chicago Tribune        & 2.74 & Newsmax          & 9.36 \\  
TPM           & 4.78 & Politico & 2.72 & WTOP-FM   & 1.17 & Laredo Morning Times   & 2.52 & Daily Caller     & 4.52 \\  
Daily Beast   & 4.11 & CBS      & 2.69 & CBS Local & 1.07 & Boston Herald          & 2.40 & Free Beacon      & 3.31 \\  
AlterNet      & 3.61 & NBC News & 2.50 & KSL News  & 1.02 & MyNorthwest            & 2.37 & The Epoch Times  & 3.07 \\  
NY Magazine   & 3.40 & NY Times & 2.46 & KTVQ-TV   & 0.95 & Daily Herald           & 2.25 & WorldNetDaily    & 2.86 \\  
Vox           & 2.69 & Newsweek & 2.03 & WFMZ-TV   & 0.95 & The Spokesman-Review   & 2.22 & TheBlaze         & 2.76 \\  
Daily Kos     & 2.68 & LA Times & 1.74 & WTHR-TV   & 0.94 & Albuquerque Journal    & 2.14 & CNSNews          & 2.64 \\  
\bottomrule
\end{tabular}
\caption{{
Statistics on outlets.
Frequency and website names of the ten top outlets in each category. 
$n$: number of outlets in each category; 
$k$: number of quotes in each category.
}
}

\label{tab:outlet_frequency_breakdown}
 \end{center}
\end{table*}

\mycomment{

\begin{table}[tb]
\footnotesize
\begin{center}
\begin{tabular}{lrlr}\\\toprule
 \textbf{Quotative} & \textbf{Percentage} & \textbf{Speaker} & \textbf{Percentage}  \\
\midrule
    say& 72.57 & Donald Trump& 15.63  \\ 
    tell& 8.82  &Barack Obama& 3.50  \\ 
    write&  3.82 & Hillary Clinton&  1.44\\ 
    tweet&  2.97 & Nancy Pelosi&  1.35\\  
    add&  2.56 & Bernie Sanders&  1.24\\  
    ask&  1.29 & Joe Biden& 1.23\\  
    continue& 0.72 & Mitch McConnell& 1.04 \\  
    respond& 0.59 & Chuck Schumer& 0.90 \\  
    declare& 0.59 & Lindsey Graham& 0.87 \\  
    state& 0.45 & Elizabeth Warren& 0.86\\  
\bottomrule
\end{tabular}
\caption{\textbf{Final dataset statistics.} Frequencies of the ten most frequent quotatives and speakers in the dataset after filtering. Among the listed quotatives, only ``declare'' is nonobjective. }

\label{tab:most_frenquent_quotative}
 \end{center}
\end{table}
}

\vspace{0.5mm}
\noindent
\textbf{Step 3: removing unsuitable outlets.}
We obtain a list of media bias ratings from \url{mediabiasfactcheck.com} (hereinafter MB/FC) and classify outlets into five categories based on the bias rating: left, left-center, least-biased, right-center, and right. {We refer to left-center, least-biased, and right-center outlets as centrist outlets in the following}. We remove quotes from outlets without a bias rating, from outlets that are not from the U.S. (also according to MB/FC data), and from outlets that have very few quotes (which may suggest data quality issues), only keeping outlets with more than 20 quotes over a period of 12 months.
After this step, around 38.0\% of the original data remains, all from relevant U.S. media outlets with human-verified bias ratings. Manual inspection of the removed data confirms that the removed outlets are predominantly non-news websites, small local newspapers, radio stations, and non-U.S news outlets.

\vspace{0.5mm}
\noindent
\textbf{Step 4: creating dictionaries of common quotatives.} 
\label{sec:verb_list}
Inspired by \citet{lee2017verb} as well as the recommendations laid out in \citet{reutershandbook} and \citet{apstylebook}, we define quotatives as \textit{objective} if they refer to the direct speech action and do not involve any subjective judgment of the action (e.g., like ``say'' and ``tell''); and as \textit{nonobjective} if they refer to some additional action or conduct and with subjective judgments (such as ``boasted,'' ``rasped,'' ``taunted,'' or ``hailed'').  To optimize for precision, we exclude common verbs with many non-quotative senses, such as ``go.'' 
Using this definition, we manually annotate the most frequent 99.5\% of quotatives overall and the 98.0\% of the most frequent quotatives per month. {We consulted a professional journalist throughout this process, who suggested that the verbs ``opine,'' ``pen,'' and ``utter'' are only sometimes used nonobjectively. Since it would be infeasible to create a separate category just for these verbs, we excluded them. 
In the end, we curated a list of 32 objective and 152 nonobjective verbs (see Appendices \ref{sec:listofverb}).} 
We use this list to remove rare verbs (i.e., those not on the list), obtaining a final dataset with 6.7M quotes (35.1\% of the original data) from 14,031 politicians in 989 outlets. 

\vspace{0.5mm}
\noindent
\textbf{Data summary:} 
Table~\ref{tab:most_frenquent_quotative} summarizes the most frequent quotatives and speakers in the final dataset.
Consistent with prior literature~\cite{gidengil2003talking}, ``say'' and ``tell'' are the most commonly used quotatives (over 80\% of the time), and Twitter is a common source for quotes~\cite{bane2019tweeting}. We find minor differences in the coverage of each speaker across outlet categories, with Donald Trump being the most quoted speaker.
Table~\ref{tab:outlet_frequency_breakdown} depicts the number of outlets per category and the fraction of quotes belonging to the ten top outlets in each category.
No single outlet dominates an entire outlet category, but the distribution is more concentrated in the left and right categories, given the relatively smaller number of outlets in these categories. 

\subsection{Validation Error Analysis}
To validate our verb extraction approach, we divide our data based on the detected verb type (objective and nonobjective) and manually annotate 100 randomly sampled quotes per year per verb type; the results are shown in Table~\ref{tab:quotative_extraction_validation}.
Our method achieves high accuracy in objective and nonobjective verb extraction, and this accuracy is stable over time, with a combined accuracy of over 90\% across all years.

Before removing quotatives that are not contained in our dictionary of common quotatives (see \textbf{Step 4}), we also analyze the instances in which quotatives with rare verbs were retrieved by our model. We find that in many cases:
{1)} the actual quotative is not a verb (e.g., ``according to'' is a common quotative phrase);
{2)} the quote is a mixed quote without quotative; or
{3)} the method is applied to incomplete or noisy data and identifies an incorrect quotative.

\begin{table}[tb]
\footnotesize

\begin{center}
\begin{tabular}{lrrr}\\
\toprule
  \textbf{Year} & \textbf{Objective} (\%) &    \textbf{Nonobjective} (\%) & \textbf{Combined} (\%) \\
  \midrule
    2013&  95 & 87 & 91\\ 
    2014& 98& 86 & 92\\
    2015& 98& 83 & 90.5\\ 
    2016& 95& 92 & 93.5\\  
    2017& 96& 90& 93 \\  
    2018& 96& 90 & 93 \\  
    2019& 98& 86 & 92 \\  
    2020& 98& 91 & 94.5 \\  

\bottomrule
\end{tabular}
\caption{Validation of the proposed quotative extraction method. We take 100 random quotes from each year that our dataset covers, manually annotate the correct quotative, and show the accuracy. Note that the incorrect cases include those in which a quotative cannot be identified. This is most often because no clear quotative is present (due to mixed quotes or because the article text is incomplete).}
\label{tab:quotative_extraction_validation}
 \end{center}
\end{table}

\subsection{Regression Model }

Throughout the results section, we fit linear probability models (LPMs) specified as 
\begin{equation}
    y_q =  \mathbf{X_q}\pmb{\beta} + \epsilon_q
\end{equation}
using ordinary least squares, where $y_q$ is a binary variable indicating whether a specific quote is nonobjective (1) or objective (0), $\mathbf{X_q}$ is an array with explanatory variables associated with quote $q$, $\pmb{\beta}$ is the array of coefficients we estimate, and $\epsilon_q$ is the error term.
Coefficients obtained in this fashion are unbiased and robust (if the predicted probabilities lie between 0 and 1~\cite{horrace2003new}, which happens for all our analyses). We estimate standard errors and $t$-statistics with cluster robust standard errors (at the outlet level), accounting for autocorrelation between quotes as well as heteroskedasticity~\cite{cameron2015practitioner}. 
More details about linear probability models can be found in \citet{wooldridge2010econometric}.

Recent work has often favored LPM instead of logistic regression or other non-linear models due to the ease of interpretation and of incorporating fixed effects~\cite{gomila2021logistic}, e.g., see \citet{dai2021behavioural}. In our case, we use the LPM since our main purpose is to approximate the partial effects of the explanatory variables~\cite{wooldridge2010econometric}. We report all results in this paper with significance level $\alpha = 0.05$.

\section{Results}\label{sec:results}

\begin{figure}
    \centering
    \includegraphics[width=0.9\linewidth]{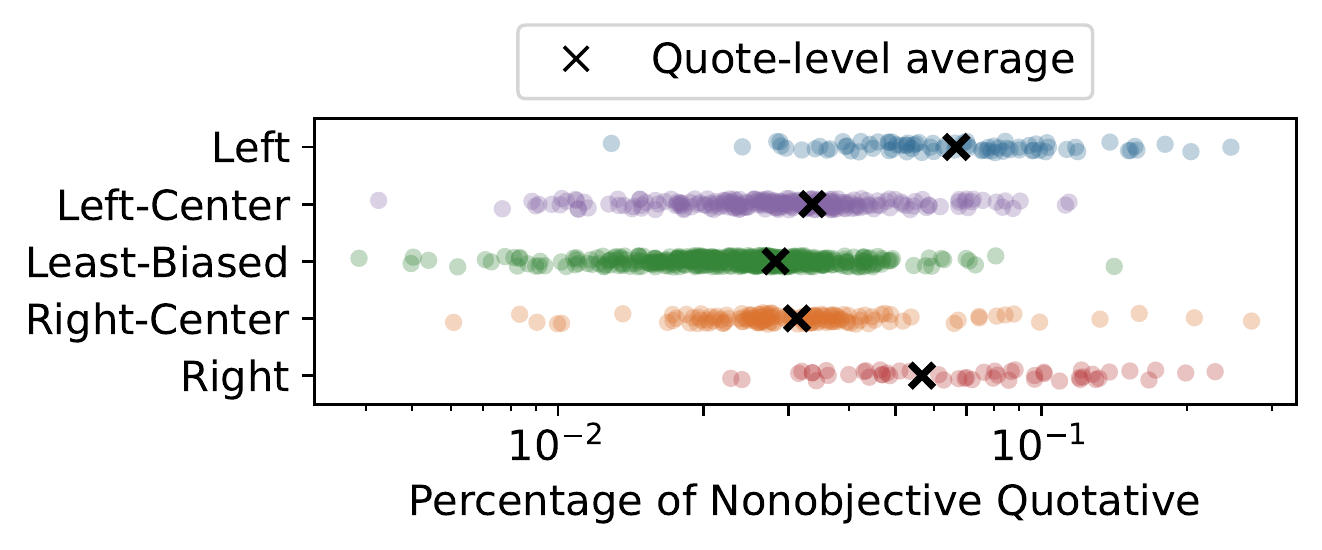}
    \caption{Usage of nonobjective quotatives across outlets of different political leaning. For each media bias category (on the $y$-axis), we depict the usage of nonobjective quotatives per outlet (each represented by a circle $\circ$) and the overall average usage pooled across outlets ($\times$). Note that the $x$-axis is on a logarithmic scale. Pairwise differences between averages are statistically significant under the Wilcoxon Rank-Sum Test with Bonferroni correction.}
    \label{fig:avg}
\end{figure}

\begin{figure}[tb]
    \begin{center}
    \includegraphics[width=7.1cm, height=2.9cm]{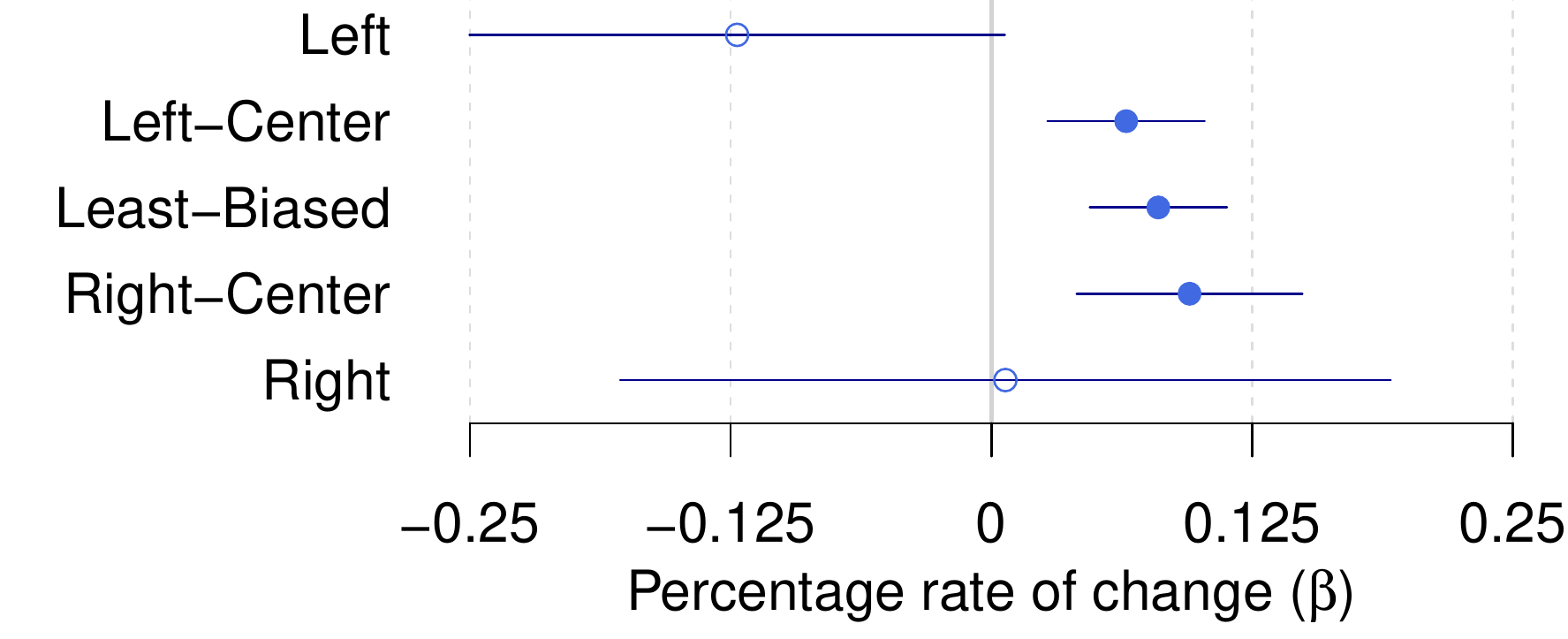} 
    \end{center}
    \caption{Percentage yearly rate of change of nonobjective quotative usage for each outlet category.  A solid circle denotes a significant effect ($p<0.05$), and a hollow circle denotes an insignificant effect. Note that the reported trends correspond to the estimated $\beta$ coefficients in Eq.~\eqref{eq1}.}
    \label{tab:regression_1}
\end{figure}

\begin{figure}[!bt]
    \begin{center}
    \includegraphics[width=0.9\linewidth]{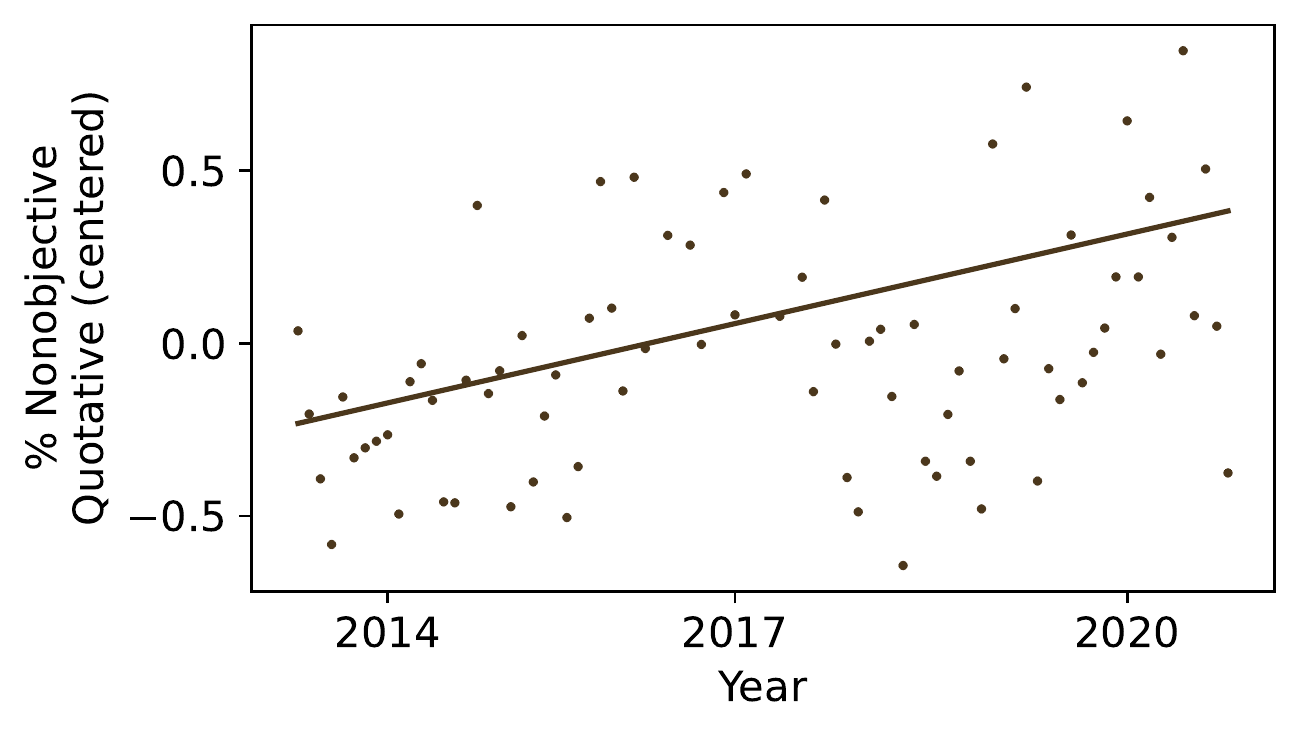} 
    \end{center}
    \caption{Percentage yearly rate of change of nonobjective quotative usage for all outlets combined. We show the percentage of nonobjective quotatives after performing centering and  plot a regression line showing the $\beta$ coefficients estimated in our fixed effects model.}
    \label{fig:trends_overall}
\end{figure}

\begin{figure*}
    \centering
    \includegraphics[width=\linewidth]{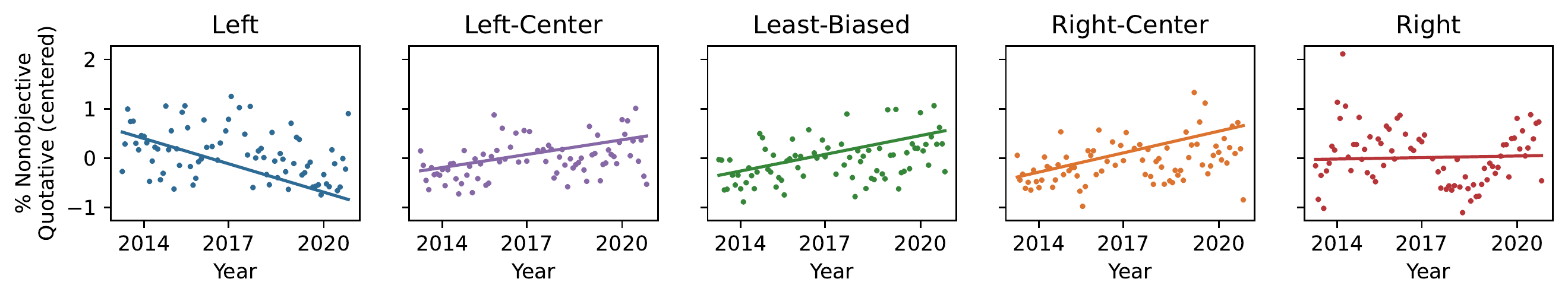}
    \caption{{Trends in the usage of nonobjective quotatives across outlets of different political leaning.} For each media bias category (one per column), we show the percentage of nonobjective quotatives after performing outlet-level centering and plot the regression line showing the $\beta$ coefficients estimated in our fixed effects model. }
    \label{fig:trends}
\end{figure*}

\subsection{RQ1: How Has the Usage of Nonobjective Quotatives Evolved?}
Across the study period, we find that the usage of nonobjective quotatives produces a sensible ordering of the media bias categories considered, with the more partisan outlets using the most nonobjective quotatives and the less partisan outlets using the least. 
We depict this order in Fig.~\ref{fig:avg}, where each circle ($\circ$) represents the average usage of nonobjective quotatives in one of the outlets considered, and crosses ($\times$) indicate the average usage pooled across each media bias category. 
When quoting politicians, U.S. least-biased outlets use nonobjective quotatives the least, followed by the left-center and right-center outlets, and finally, right and left outlets. 
Outlets considered more partisan by MB/FC used nonobjective quotatives more. Pairwise differences between averages are statistically significant under the Wilcoxon Rank-Sum Test with Bonferroni correction.

To study how the usage of nonobjective quotatives evolved, we use a fixed effects linear probability model.
For each quote $q$, let $o[q]$ be the outlet in which $q$ was reported, $b[q]$ be the bias category of the outlet, and $t[q]$ be the time when it was in reported in months relative  to the starting period of our dataset (May 2013). 
We define the model as 
\begin{equation}
\label{eq1}
    y_{q} = 
    \alpha_{o[q]} + 
    \gamma_{b[q]} + \beta_{b[q]}  t[q] + \epsilon_{q},
\end{equation}
where the dependent variable $y_{q}$ equals 1 if the verb used in the quote $q$ is nonobjective and 0 otherwise, $\alpha_{o[q]}$ is an outlet-level intercept, $\gamma_{
b[q]}$ is a category-level intercept, and $\beta_{b[q]}$ is a category-level trend in the usage of nonobjective quotatives -- the effect we are interested in estimating.%
Since we are modeling time-series data (one per outlet), autocorrelation may shrink the confidence intervals of model estimates (see \citet{bertrand2004much} for details). To address this, we estimate the model using cluster robust standard errors, clustering on the outlet level~\cite{cameron2015practitioner}.

\begin{table}[tp]
\footnotesize
\begin{center}

\begin{tabular}{@{}llll@{}}\\\toprule
 Quotative & Percentage Change & Quotative & Odds Ratio\\
 \midrule
    say& -10.18$\downarrow$ & tweet& 1 $\rightarrow$ 17.04\\
    tweet& 4.157$\uparrow$& \textit{falter} & 1 $\rightarrow$ 11.88\\  
    tell&  1.843$\uparrow$&  caption & 1 $\rightarrow$ 10.86  \\ 
    write&  1.555$\uparrow$ & \textit{restate} & 1 $\rightarrow$ 6.772  \\  
    add&  1.020$\uparrow$ & \textit{remark} & 1 $\rightarrow$ 5.614\\  
    respond & 0.3538$\uparrow$ &  \textit{punctuate} & 1 $\rightarrow$ 4.855\\  
    continue& 0.3454$\uparrow$ &      \textit{blurt} & 4.891 $\rightarrow$ 1 \\
    \textit{declare} & 0.2325$\uparrow$ &   \textit{disclose} & 4.911 $\rightarrow$ 1 \\  
    \textit{remark}& 0.2159$\uparrow$&  \textit{enthuse} & 6.335 $\rightarrow$ 1 \\
    \textit{claim}& 0.2028$\uparrow$ &  \textit{exult} & 15.65 $\rightarrow$ 1\\

\bottomrule
\end{tabular}
\caption{
{
Changes in quotatives used. 
We report the most changed quotatives between our dataset's first and last 12 months in absolute change and odds ratio. \textit{Italic} highlighting denotes nonobjective quotatives.
}}
\label{tab:most_quotative_change}
 \end{center}
\end{table}

We depict the estimated trends in nonobjective quotative usage in Fig.~\ref{tab:regression_1} (i.e., the estimated $\beta_{b[q]}$ in Equation \eqref{eq1}).
Although the least-biased outlets used nonobjective quotatives less on average (Fig.~\ref{fig:avg}), we find that their usage of nonobjective quotatives increases over time. We estimate that least-biased outlets increase their usage of nonobjective quotatives by 0.08\% per year and that right-center and left-center outlets increase their usage by 0.10\% and 0.06\% per year, respectively. 
{If we compare the level of nonobjective quotative usage from 2013 to 2020 (beginning and end of our study)}, these changes translate to relative increases of  19.9\% for least-biased outlets,  21.3\% for right-center and 13.6\% for left-center outlets.
In contrast, left outlets experienced a statistically insignificant decrease in their usage of nonobjective quotatives by 0.12\% per year, and right outlets experienced a smaller, statistically insignificant increase in usage of nonobjective quotatives (of roughly 0.01\%). 
We also experimented with an added seasonality factor shared across all outlets and politicians, either on a monthly or yearly basis. We exclude these results here since 1) accounting for seasonality does not substantially change the results, 2) effect sizes decrease only slightly when seasonality is considered, and 3) with only two four-year cycles, there is insufficient data for a robust analysis of seasonality.

We further illustrate the results obtained in the fixed effects model in  Figure~\ref{fig:trends_overall} and~\ref{fig:trends}. In Figure~\ref{fig:trends_overall}, we center the overall quotative usage around 0 and plot the month-level nonobjective quotative usage, along with a regression line capturing the trend. The increase in nonobjective quotative usage across all outlets aggregated in percentage per year (i.e., the slope) is 0.079\% ($p < 0.001$). In other words, we find that the overall rate of nonobjective quotative usage among all outlets is increasing, thereby supporting the argument that changes in the usage of quotatives indicate a steady decline in objectivity in U.S. political news.

In Figure~\ref{fig:trends}, we plot at the outlet category level: we center each outlet time series around 0 and then report the month-level (demeaned) usage of nonobjective quotatives per outlet category, along with a regression line capturing the trend in each time series.
Here, we observe that the usage of nonobjective quotatives increases for centrist outlets, decreases for outlets on the left, and only slightly increases for outlets on the right, although the two latter results were not statistically significant according to the model. In other words, the overall trend in decreasing objectivity can be explained by the trends in centrist outlets.

Another way to understand the change in quotative usage is to consider the extremes. We compare how quotative usage changes between the first 12 months (May 2013 - April 2014) and the last 12 months (May 2019 - April 2020) of our dataset.
In Table \ref{tab:most_quotative_change}, we report the quotatives that experienced the largest changes in terms of absolute percentage points (on the left) and odds ratio (on the right). We find that the usage of the quotative ``say,'' typically considered the gold standard of quotatives, fell by more than 10\% percentage points. At the same time, we see an increase in other objective quotatives (e.g., tell), but this increase does not account for the entire ten percentage points. Lower in the list, we see that nonobjective quotatives like ``claim,'' ``remark,'' and ``declare'' are used more often.
Finally, we highlight that quotatives reveal changes in where journalist source their quotes, with both ``tweet'' and ``caption'' (usually employed when the speaker uploads a picture or video on social media) experiencing large relative increases in usage.

\subsection{RQ2: How Do News Outlets Use Nonobjective Quotatives
When Covering Politicians of Different Parties? } \label{subsection:interaction_leaning_party}

Next, we investigate whether the outlets are biased in their quotative usage when they cover politicians from ideologically similar vs.\ opposing political parties. 

\vspace{1.5mm}
\noindent
\textbf{Quotative bias across outlet categories.}
For each quote $q$, let $p[q]$ be the party of the politician who uttered the quote. Keeping with the notation in Eq.~\eqref{eq1}, we again use a fixed effects linear probability model
\begin{equation}
\label{eq2}
    y_q =  \gamma_{b[q]} + \eta_{p[q]} + \sigma_{b[q],p[q]} + \epsilon_{q},
\end{equation}
where the dependent variable $y_q$ equals 1 if the quotative used in the quote $q$ is nonobjective and 0 otherwise, 
$\gamma_{b[q]}$ is a category-level intercept,
$\eta_{p[q]}$ is a party-level intercept, and 
$\sigma_{b[q],p[q]}$  captures the interaction between pairs of outlet bias category ($b[q]$) and speaker party ($p[q]$). We emphasize that outlet and outlet category are outlet-level attributes, while the political party is a politician-level attribute.
We again cluster standard errors on the outlet level to address autocorrelation.
Note that here, we are particularly interested in the contrasts between different combinations of outlet categories and speaker parties, e.g., the difference between how left outlets quote Democratic and Republican speakers (in the model $\sigma_{\text{left}, \text{democratic}} -  \sigma_{\text{left}, \text{republican}}$). 

 \begin{figure}
    \begin{center}
    \includegraphics[width=8.2cm, height=3.5cm]{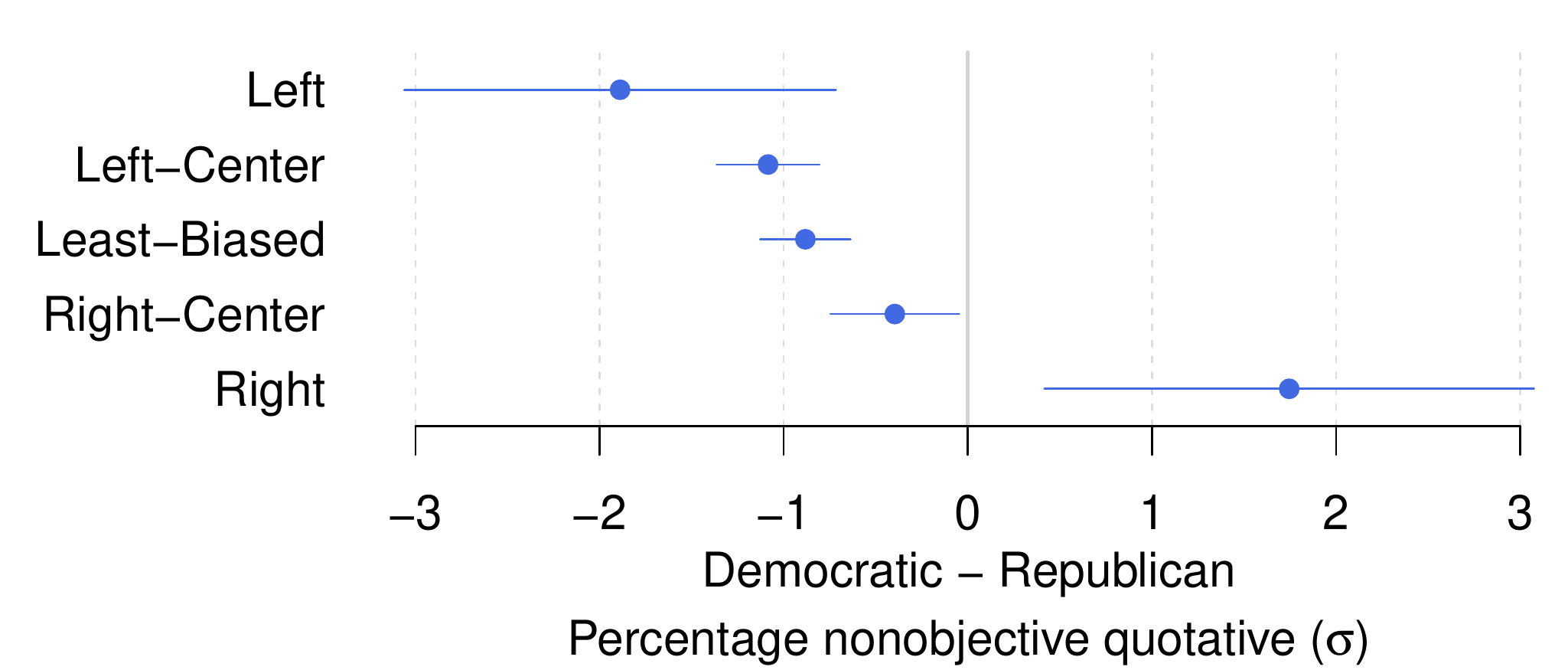} 
    \end{center}
    \caption{
    Differences in nonobjective quotative usage between Democratic and Republican speakers. All estimates are significant.
    Reported differences correspond to the contrasts $\sigma_{\text{democratic}} -  \sigma_{\text{republican}}$ in Eq.~\eqref{eq2} {(in percentage points)}.
    }
    \label{fig:regression_2a}
\end{figure}

For each media bias category, we show the estimated percentage difference in the usage of nonobjective quotatives for Democratic and Republican speakers in Figure~\ref{fig:regression_2a}.
For every outlet category, there is a significant difference in quotative usage between Democratic and Republican speakers. 
Notably, this difference is nearly 2\%, around a third of the overall nonobjective quotative usage, for both left and right media outlets, which use more nonobjective quotatives when referring to politicians from opposing political parties.
For centrist outlets, we see a Democratic bias in the usage of quotatives, with Republicans being quoted with nonobjective quotatives around 1\% more for least-biased and left-center outlets and nearly 0.5\% more for right-center outlets.

\vspace{1.5mm}
\noindent
\textbf{Matched analysis.} 
A possible explanation for what we observe in Figure~\ref{fig:regression_2a} is that outlets of different political leanings cover different quotes~\cite{tan2018you} and that these quotes lend themselves more or less to being attributed to speakers through nonobjective quotatives.
Even if this were the case, one could still make a case against media bias, as journalistic textbooks and guidelines instruct the usage of objective quotatives regardless of the quote~\cite{mencher1997news,brooks2007news,reutershandbook,apstylebook,rich2015writing}. 
Nevertheless, we entertain this hypothesis by performing a matched analysis. 
Specifically, we identify quotes covered by both the left and right media outlets,  merging left/left-center and right/right-center outlets for ease of comparison (we refer to these merged bias categories as left and right ``combined'').
Using only this subset of quotes ($n=1.02$M, 15.13\% of all quotes in our data), we fit the fixed effects model defined in Eq.~\eqref{eq2}.

\begin{figure}
    \begin{center}
    \includegraphics[width=8.2cm,height=2.5cm]{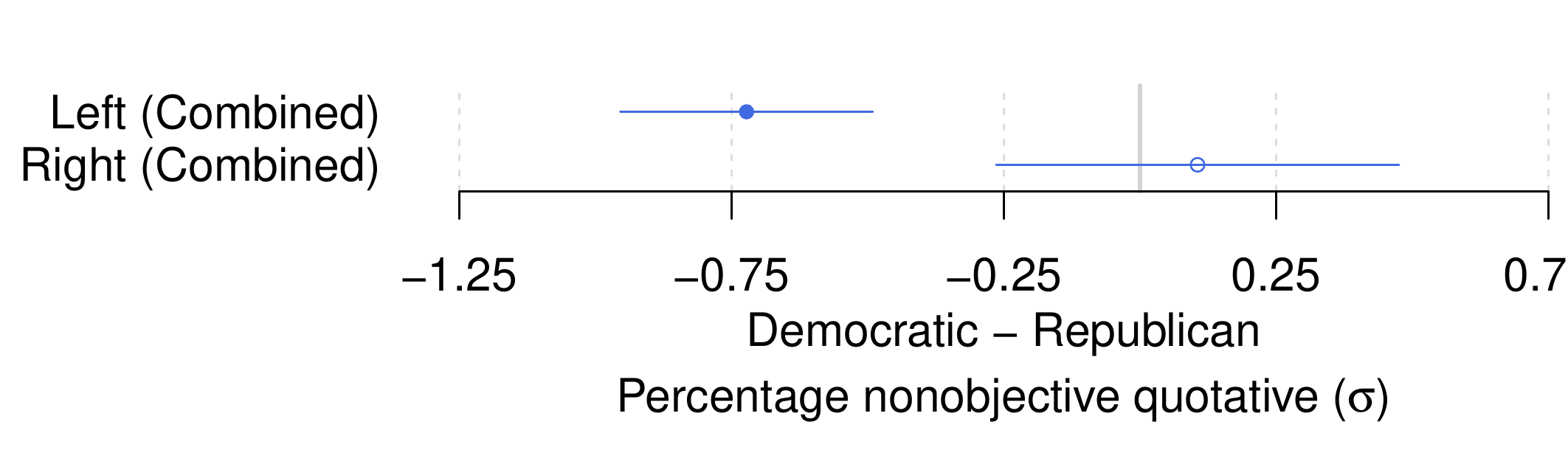} 
    \end{center}
    \caption{Differences in nonobjective quotative usage between Democratic and Republican speakers for matched quotes. A solid circle denotes a significant effect ($p < 0.05$), and a hollow circle denotes an insignificant effect.
    Reported differences correspond to the contrasts $\sigma_{\text{democratic}} - \sigma_{\text{republican}}$ in Eq.~\eqref{eq2} {(in percentage points)}. 
     }
     \label{tab:regression_3a}
\end{figure}

For the two collapsed media bias categories and considering only matched quotes, we show the estimated difference in usage of nonobjective quotative for Democratic and Republican speakers in Figure \ref{tab:regression_3a}.
For right/right-center outlets, we find a small non-significant positive difference (0.01\%).
Note that in the non-matched scenario, in Figure~\ref{fig:regression_2a}, these two types of outlets behave differently -- right outlets use more nonobjective quotatives when quoting democrats, whereas right-center outlets do so when quoting republicans. 
Since we aggregate right and right-center, it may be that these heterogeneous effects cancel each other.
For left/left-center outlets, we find a significant difference of around -0.75\% in the usage of nonobjective quotatives. This suggests that quote selection alone cannot explain the quotative bias previously observed and that quotative selection forms an additional source of bias on top of quote selection.

\vspace{1.5mm}
\noindent
\textbf{Trends in quotative bias.} 
Finally, we investigate if quotative bias has evolved during the study period, using a fixed effects linear probability model:
\begin{equation}
\label{eq3}
    y_q =  \alpha_{o[q]} + \gamma_{b[q]} + \eta_{p[q]} + \lambda_{b[q],p[q]} t[q] + \epsilon_{q},
\end{equation}
where the dependent variable $y_q$ equals 1 if the quotative used in the quote $q$ is nonobjective and 0 otherwise, 
$\alpha_{o[q]}$,
$\gamma_{b[q]}$, and
$\eta_{p[q]}$ are outlet, category, and party-level intercepts, 
and 
$\lambda_{b[q],p[q]}$ is the trend in the usage of nonobjective quotatives for each party/bias category combination. 

\begin{figure}[t]
    \begin{center}
    \includegraphics[width=8.2cm, height=3.5cm]{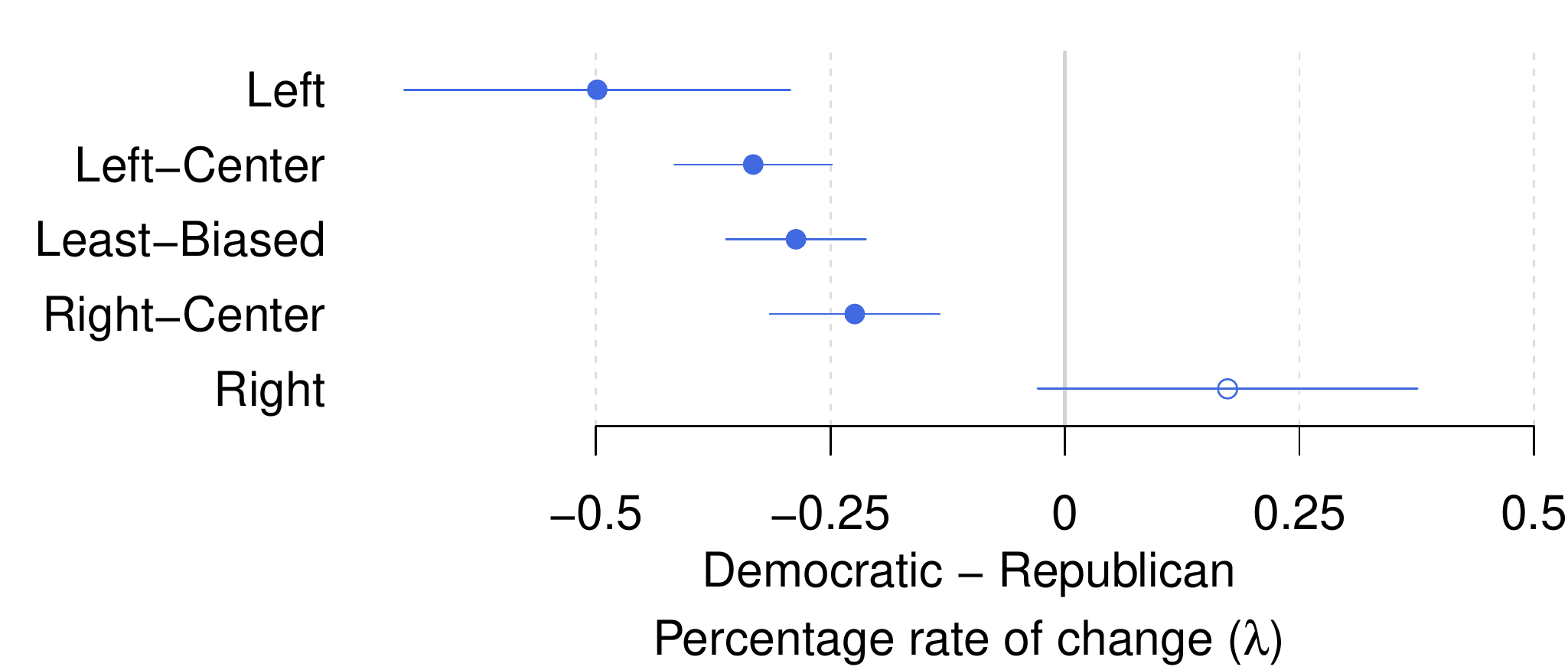} 
    \end{center} 
     \caption{
   Differences in percentage yearly rate of change of nonobjective quotative usage between Democratic and Republican speakers for each media outlet category.
   A solid circle denotes a significant effect ($p < 0.05$). Reported trends correspond to the contrasts $\lambda_{\text{democratic}} - \lambda_{\text{republican}}$ in Eq.~\eqref{eq3} {(in percentage points).}
   }

   \label{tab:regression_4}
\end{figure}

For each media bias category, we depict the difference in the trends of nonobjective quotative usage for Democrats and Republicans in Figure~\ref{tab:regression_4}. 
For left and centrist outlets, the gap between how nonobjective quotatives are used to quote Democrats and Republicans is increasing in the study period. These increases are statistically significant and substantial compared to the existing level of quotative bias observed in our data. For example, the estimated contrast of the trend is around 0.33\% for left-center outlets, and the existing quotative bias is 1.08\%. Thus, the annual relative increase of quotative bias is above 30\%. Left outlets exhibit the most increase in quotative bias in absolute terms, at 0.5\% per year.
For right outlets, the difference in trend leans Republican, but the effect is not statistically significant.

\section{Discussion}\label{sec:discussion}
In this work, we analyzed quotatives to study political journalism, answering the following questions: How has the usage of nonobjective quotatives evolved in U.S. political journalism (\textbf{RQ1})? 
How do news outlets use nonobjective quotatives when covering politicians of different parties (\textbf{RQ2})? 
To answer these questions, we proposed a method to identify quotatives for direct quotes using dependency parsing. We then extracted quotatives from a large dataset of speaker-attributed quotes, resulting in over 6.7 million quotes over eight years, from 2013 to 2020. By counting the usage of objective and nonobjective quotatives, we analyzed the static and dynamic trends of quotative usage.

We find that the more partisan outlets use more nonobjective quotatives (Figure~\ref{fig:avg}). 
However, during the study period, centrist outlets (classified as least-biased, left-center, and right-center by MB/FC) experienced a significant increase in the usage of nonobjective quotatives, suggesting that they may be ``catching up'' to the more biased outlets (Figure~\ref{fig:trends}). 
We further observe that outlets tend to use more nonobjective quotatives when covering politicians of the opposing ideology, thereby exhibiting ``quotative bias'' (Figure~\ref{fig:regression_2a}). Even when we control for quotes by matching outlets on the quote level, we find that this bias still exists for left and left-center outlets (Figure~\ref{tab:regression_3a}).
Finally, we find a rapid increase in quotative bias for most outlet categories over time, which may indicate that U.S. political news is becoming increasingly polarized (Figure~\ref{tab:regression_4}).

These findings suggest that two simultaneous processes are at play: outlets are adopting more nonobjective quotatives overall \emph{and} the usage of nonobjective quotatives is increasingly ``mediated'' by the party affiliation of quoted politicians. Both processes indicate a measurable decrease in journalistic objectivity.
While detecting bias often requires some level of human judgment to determine neutrality, and while it is debatable how a neutral or balanced view can be presented in any specific context, quotative usage can be regarded as an easily quantifiable form of bias due to its prominence within journalism. There are clear and established rules for the usage of quotes on which journalists have historically agreed, as is evident from textbooks~\cite{brooks2007news,mencher1997news,rich2015writing} and editorial guidelines~\cite{reutershandbook,apstylebook}.
Although objective journalism is a 20th-century invention and could be considered an anomaly throughout history, it is regarded as central to today's democratic process. 
In this context, our results indicate a decrease in the level of objective quotative usage in U.S. political news coverage, which can be seen as a devolution of journalism as a profession. 

An interesting question is to which extent Donald Trump, the most quoted speaker in our dataset, influences the findings observed in this paper. To investigate this, we re-run all fixed effects models in the paper in a filtered dataset from which we removed Trump's quotes (see Appendices \ref{app:trump}). We find that, without Trump, there is no significant increase in the overall usage of nonobjective quotatives. However, we still observe quotative bias in all outlet categories except left-center outlets.

The ways in which the observed increase in nonobjective quotatives relates to broader trends in U.S. politics and the news ecosystem remains unclear. 
On the one hand, the observed trend may merely reflect the reality of the news business. 
As newspapers struggle to retain subscribers and attract clicks~\cite{thurman2019algorithms}, outlets (including the least-biased ones) may have succumbed to nonobjective quotatives as they adapt to the fast-paced style of Web-first publishing and try to produce engaging content.
Alternatively, journalists themselves may be subject to trends of increasing polarization in the general public~\cite{abramowitz2008polarization}, becoming more prone to Freudian slips when reporting the speech of politicians they (dis)like. 
On the other hand, the increase in quotative bias may influence people's opinions about politicians~\cite{cole1974powerful} or erode the reader's trust in the media outlet, as they might disagree with the opinions subtly embedded in the news piece by the writer~\cite{gunther1992biased}.

\vspace{1mm}
\noindent
\textbf{Limitations.}
We highlight three limitations of our work. 
First, while Quotebank covers a large number of outlets over time, the number of quotes from each outlet does not necessarily reflect the actually published amount of content per outlet, nor does it represent the relative popularity of each outlet. Therefore, our findings should not be interpreted as the perception that an average news consumer in the U.S. may have.
Second, since we focus on quotative verbs and do not consider adverbs, we may not capture the complete picture of quotative usage. For example, the hypothetical quotative ``say aggressively'' should be categorized as nonobjective, but we still categorize it as objective since we do not consider adverb usage. Empirically, however, this combination of an objective quotative verb and a nonobjective adverb is uncommon (as it can be considered even more unprofessional than a nonobjective quotative verb) and is unlikely to affect our results. The occurrence of non-verb quotatives is rare as well. Also, despite our best effort, a few quotatives may be misidentified or not included in our dictionary. 
While this should not affect our overall analysis, analyzing these rare quotatives might yield additional insights. 
Third, while we carefully removed most indirect and mixed quotes from our dataset, some are expected to remain in the data. 
Thus, we cannot completely eliminate the effect of indirect quotes on the obtained results. 

\vspace{1mm}
\noindent
\textbf{Future Work.} 
Future work could use and extend our methodology to investigate trends in nonobjective quotative usage after the Trump presidency or in other countries and languages. 
Further, an even more comprehensive investigation of the landscape of quotative usage in political journalism could be obtained by extending our methodology to include indirect and mixed quotes and/or considering nonverb phrases and adverbs as quotatives.
Last, future work could examine if nonobjective quotatives reflect changes in polarization or the media ecosystem and to what extent they impact readers' opinions of politicians and news outlets.

\begin{figure}[tb]
\begin{subfigure}[b]{\linewidth}
\centering
\includegraphics[width=7cm, height=2.9cm]{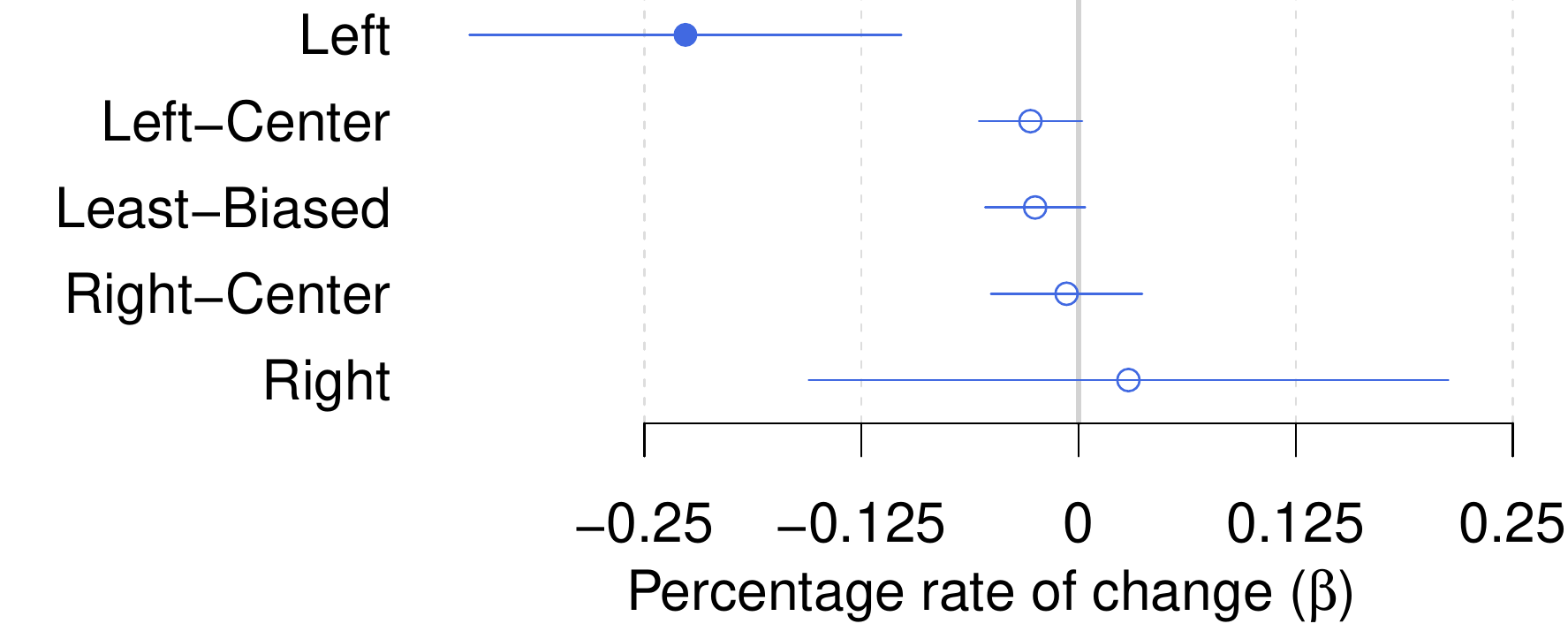} 
\caption{Same as  Figure \ref{tab:regression_1}.}
\end{subfigure}

\begin{subfigure}[b]{\linewidth}
\centering
\includegraphics[width=8.2cm, height=3.5cm]{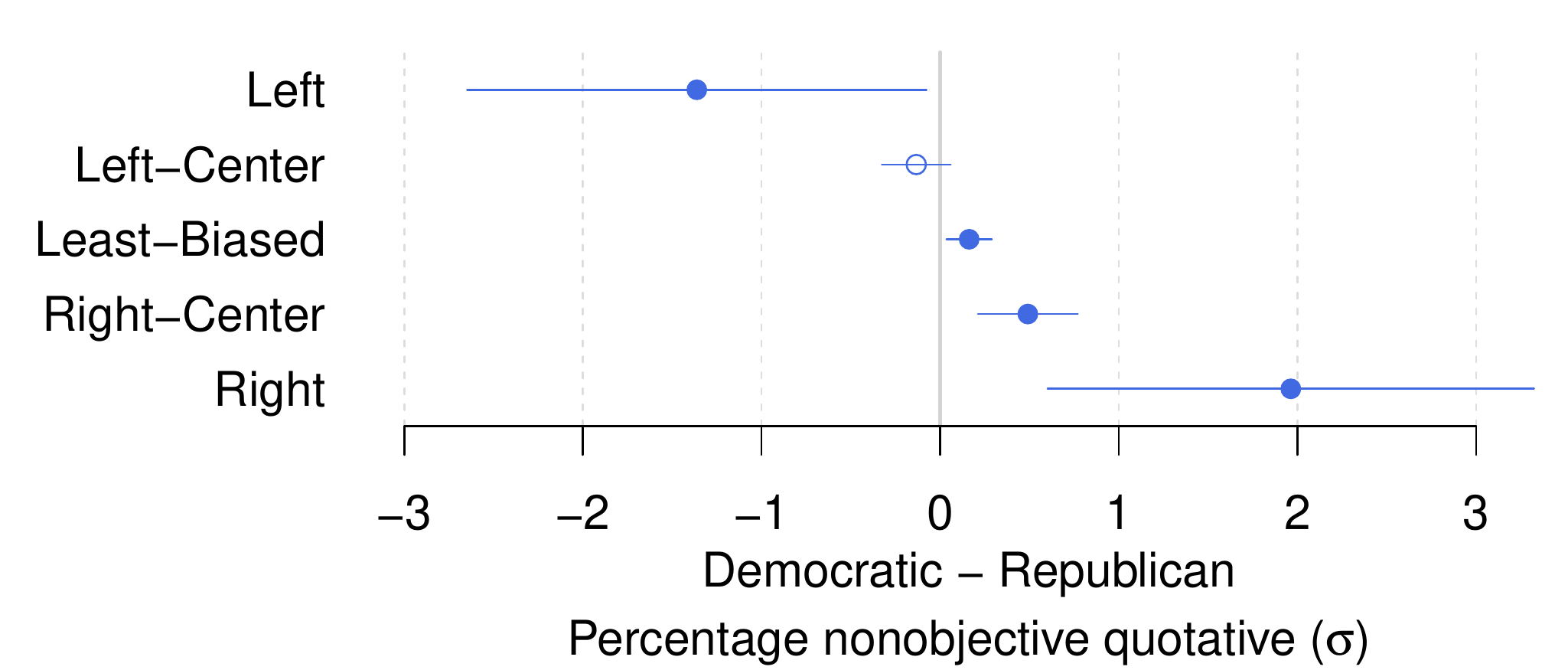} 
\caption{Same as  Figure \ref{fig:regression_2a}.}
\end{subfigure}

\begin{subfigure}[b]{\linewidth}
\centering
\includegraphics[width=8.2cm, height=3.5cm]{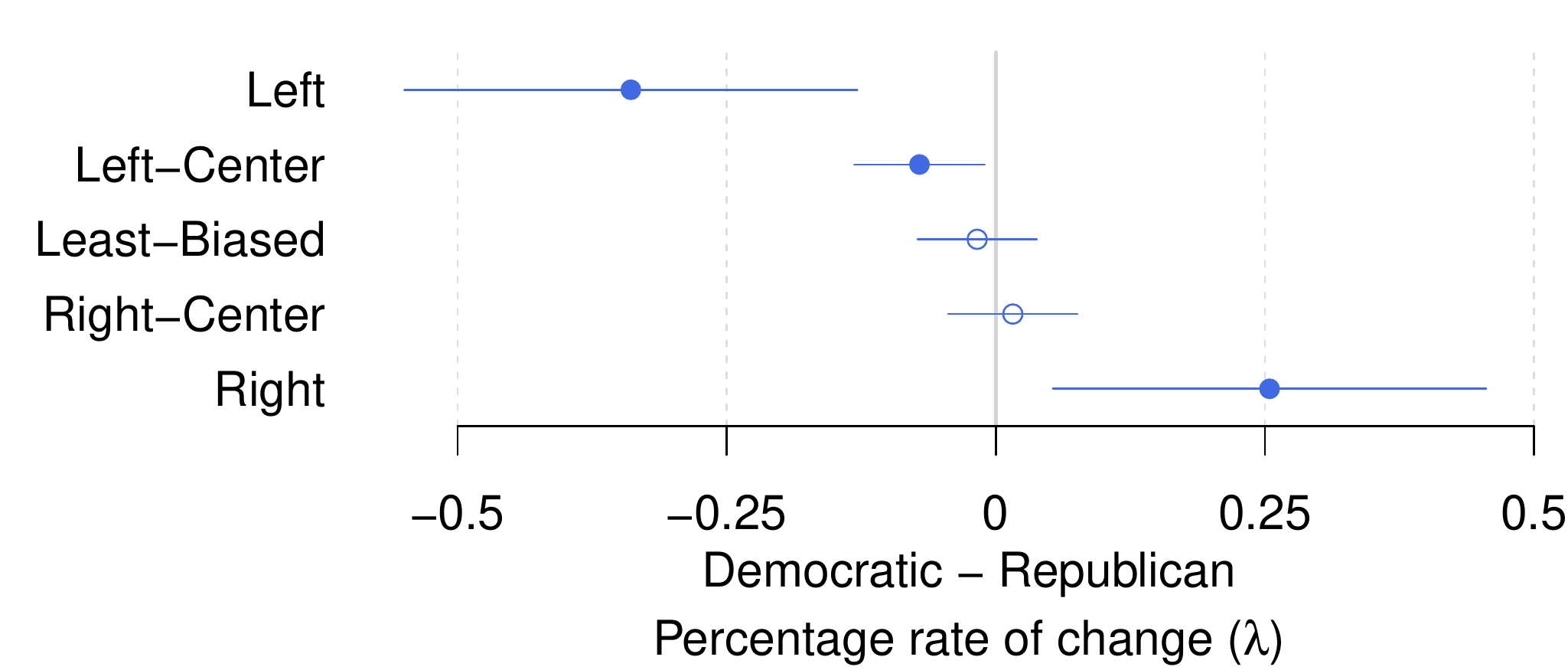} 
\caption{Same as  Figure \ref{tab:regression_4}.}
\end{subfigure}

\caption{
These figures correspond to the same analysis from Figures \ref{tab:regression_1},
\ref{fig:regression_2a}, and
\ref{tab:regression_4} in the main paper but without any quotes from Donald Trump.}
\label{supfig:trump}
\end{figure}

\appendix
\section{Appendices}
\subsection{List of Objective and Nonobjective Quotatives} \label{sec:listofverb}
\textbf{Objective Quotative}: {say tell write tweet add ask continue respond state explain note read reply quote announce recall conclude post begin describe answer cite testify recount close email summarize finish caption inform accord preface }

\noindent
\textbf{Nonobjective Quotative}: {declare warn claim argue insist remark joke suggest acknowledge urge promise comment assert quip proclaim admit share complain vow praise stress predict charge shoot observe emphasize boast reiterate pledge remind fire counter lament shout concede caution assure retort confirm exclaim contend advise laugh blast hit indicate yell press reflect tout fume mock muse interject gush apologize brag clarify thunder challenge hail interrupt snap elaborate chide chime plead lash intone confess disagree protest crow boom tease cry scold laud hint affirm crack implore scoff bellow chuckle rail lecture smile speculate scream bemoan reassure shrug marvel rip underscore decry commend gripe object confide jab pronounce taunt instruct enthuse admonish roar chastise whine rant reminisce reaffirm concur recite disclose beam whisper deflect posit rebuke pile falter articulate deride channel sneer blurt persist grumble ratchet punctuate forecast sigh sketch exhort explode burst preach cede interrogate diagnose gloat tee shy wax mourn exult goad backpedal restate howl}

\begin{table}[ht]

\begin{subfigure}[b]{\linewidth}
\footnotesize
\centering
\caption{Coefficients shown in Fig.~\ref{tab:regression_1}.}
\begin{tabular}{lrrrr}
  \toprule
Outlet & $\beta$ & SE & $t$ & $p$ \\ 
  \midrule
  All & 0.08 & 0.02 & 4.14 & $<$0.001 \\
Left & -0.12 & 0.07 & -1.86 & 0.063 \\ 
  Left-Center & 0.06 & 0.02 & 3.36 & 0.001 \\ 
  Least-Biased & 0.08 & 0.02 & 4.77 & $<$0.001 \\ 
  Right-Center & 0.10 & 0.03 & 3.44 & 0.001 \\ 
  Right & 0.01 & 0.09 & 0.07 & 0.943 \\ 
   \bottomrule
 \label{tab:reg_coeff_1}
\end{tabular}
\end{subfigure}

\begin{subfigure}[b]{\linewidth}
\footnotesize
\centering
\caption{For coefficients shown in Fig.~\ref{fig:regression_2a} (top) and Fig.~\ref{tab:regression_3a} (bottom). }
\begin{tabular}{lcrrr}
  \toprule
Outlet & $\sigma_{\text{dem}} -  \sigma_{\text{rep}}$  & SE & $t$ & $p$ \\ 
  \midrule
Left & -1.89 & 0.37 & -5.10 & $<$0.001 \\ 
  Left-Center & -1.09 & 0.09 & -12.26 & $<$0.001 \\ 
  Least-Biased & -0.88 & 0.08 & -11.36 & $<$0.001 \\ 
  Right-Center & -0.40 & 0.11 & -3.58 & 0.013 \\ 
  Right & 1.75 & 0.42 & 4.15 & 0.001 \\ 
  \midrule
  Left (Combined)& -0.72 & 0.09 & -8.00 & $<$0.001 \\ 
  Right (Combined) & 0.10 & 0.14 & 0.73 & 0.883 \\ 
  \bottomrule
 \label{tab:reg_coeff_2}
\end{tabular}

 \end{subfigure}
\begin{subfigure}[b]{\linewidth}
\footnotesize
\centering
\caption{For coefficients shown in Fig.~\ref{tab:regression_4}. }
 \begin{tabular}{lcrrr}
  \toprule
Outlet & $\lambda_{\text{dem}}-\lambda_{\text{rep}}$  & SE & $t$ & $p$ \\ 
  \midrule
Left & -0.50 & 0.07 & -7.67 & $<$0.001 \\ 
  Left-Center & -0.33 & 0.03 & -12.50 & $<$0.001 \\ 
  Least-Biased & -0.29 & 0.02 & -12.18 & $<$0.001 \\ 
  Right-Center & -0.22 & 0.03 & -7.82 & $<$0.001 \\ 
  Right & 0.17 & 0.06 & 2.71 & 0.170 \\    \bottomrule
 \label{tab:reg_coeff_1x}
\end{tabular}
 \end{subfigure}

     \caption{
     {
     Regression result details. For key coefficients discussed throughout the paper, we present standard errors, $t$ statistics, and $p$ values.}
}

\end{table}

\subsection{Analysis of Quotes Without Donald Trump}
\label{app:trump}
In Figure~\ref{supfig:trump}, we report the coefficients of interest for the fixed effects models depicted in Equations \eqref{eq1}, \eqref{eq2}, and \eqref{eq3} on a filtered dataset containing no quotes by Donald Trump.

\section*{Ethical Statement}

This work uses publicly available data to analyze quotes from U.S. politicians to study media bias. We match politicians' names between Quotebank and Wikidata. 
Our study utilizes one identity characteristic -- the political party of the involved speakers -- at an aggregated level in order to investigate the effect of ideology in quotative usage. 
We do not perform any individual-level inferences.
Additionally, we manually validate both speaker disambiguation and quotative extraction methods to minimize the risk of identifying the wrong individuals and ensure the veracity of our findings. Given that politicians are public figures and the importance of research to better understand the language used in political journalism and its implications, we believe that our work is in line with reasonable expectations of privacy~\cite{doherty2007politicians}. 
We do not foresee potential negative societal impacts coming from this research.
On the contrary, we believe that a better understanding of our political media ecosystem is essential to improve it.
We confirm that we have read and abide by the AAAI
code of conduct.

\section*{Acknowledgements}

We would like to thank Jonathan K\"{u}lz and Marko \v{C}uljak for their help in data preprocessing and Chris Stokel-Walker for having a look at our list of quotatives.
This project was partly funded by the Swiss National Science Foundation (grant 200021\_185043), the European Union (TAILOR, grant 952215), the Microsoft Swiss Joint Research Center, and the Deutsche Forschungsgemeinschaft under the Excellence Strategy of the German federal and state governments (DFG, grant EXC-2035/1 - 390681379). We also acknowledge generous gifts from Facebook and Google supporting West’s lab.

{\small
\bibliography{bib}
}

\end{document}